\documentclass[journal]{IEEEtran}
\IEEEoverridecommandlockouts
\usepackage{times,amsmath,color,amssymb,graphicx,epsfig,cite,psfrag,subfigure,algorithm,balance}
\usepackage{amsfonts,pifont,enumerate,cases}
\usepackage{mathrsfs} 
\usepackage[table]{xcolor} 
\usepackage{verbatim} 
\usepackage{bm}
\usepackage{cuted,stfloats}
\usepackage{algorithm}
\usepackage{algorithmic}

\usepackage{longtable}
\usepackage{blindtext}
\usepackage{multirow}
\usepackage{float}
\usepackage{threeparttable}
\usepackage{makecell}
\usepackage[utf8]{inputenc}
\usepackage{url}
\usepackage{booktabs}
\usepackage{amssymb}
\usepackage{bbding}
\usepackage{pifont}
\usepackage{wasysym}
\usepackage{utfsym}
\usepackage{fontawesome}
\usepackage[algo2e,ruled,vlined,linesnumbered,lined,boxed,commentsnumbered]{algorithm2e}
\usepackage{amsmath,mathtools}
\usepackage[
    colorlinks=true,
    linkcolor=blue,
    citecolor=blue,
    urlcolor=magenta
]{hyperref}
\usepackage{array}

\begin{document}
\title{Dual-Layer Over-the-Air Federated Learning in LEO Satellite Networks: Architecture, Key Technologies and Applications}
\author{Zhendong Li, Shaojie Wang, Zhou Su, Tom H. Luan, Ruijin Sun, Ying Wang, and Wen Chen
\thanks{Zhendong Li is with the School of Information and Communication Engineering, Xi'an Jiaotong University, Xi'an 710049, China, is also with the State Key Laboratory of Integrated Services Networks and School of Telecommunications Engineering, Xi’an 710126, China (email: lizhendong@xjtu.edu.cn). Shaojie Wang is with the School of Information and Communication Engineering, Xi'an Jiaotong University, Xi'an 710049, China (email: wessel@stu.xjtu.edu.cn). Zhou Su and Tom H. Luan are with the School of Cyber Science and Engineering, Xi'an Jiaotong University, Xi'an 710049, China (email: zhousu@ieee.org; tom.luan@xjtu.edu.cn). Ruijin Sun is with the State Key Laboratory of Integrated Services Networks and School of Telecommunications Engineering, Xidian University, Xi'an 710126, China (e-mail: sunruijin@xidian.edu.cn). Ying Wang is with the State Key Laboratory of Networking and Switching Technology, Beijing University of Posts and Telecommunications, Beijing 100876, China (e-mail: wangying@bupt.edu.cn). Wen Chen is with the Department of Electronic Engineering, Shanghai Jiao Tong University, Shanghai 200240, China (e-mail: wenchen@sjtu.edu.cn). (Corresponding author: Zhou Su)}
\vspace{-1.5em}
}
\maketitle
\thispagestyle{empty}

		\maketitle
		
\begin{abstract}
Low Earth orbit (LEO) satellite networks are emerging as a pivotal infrastructure for global edge intelligence. In this context, integrating over-the-air (OTA) computation with adaptive beam hopping (BH) provides an innovative framework that seamlessly merges physical-layer analog aggregation with dynamic resource orchestration. 
This effectively overcomes the stringent bandwidth and power constraints of space platforms while extending federated learning (FL) to pervasive Internet-of-things (IoT) deployments. 
In this article, we first outline the fundamental principles of the dual-layer OTA model and introduce the adaptive BH mechanism designed for time-varying topologies. Then, we summarize the distinct advantages of this learning-centric architecture, which include decoupling aggregation latency from device density, optimizing spatio-temporal resource efficiency, and balancing data freshness with channel quality. 
Several application scenarios are explored to highlight the framework's potential across diverse vertical industries.
Furthermore, a specific case is studied to demonstrate the practical efficacy of the proposed scheduling policy. The results reveal substantial performance gains in terms of model convergence speed and data utilization for satellite-based FL systems. 
Finally, we discuss the implementation challenges and outline future research directions, aiming to provide insights for the evolution of ubiquitous non-terrestrial intelligence.

\end{abstract}
		
			
		
\section{Introduction}
 
\IEEEPARstart{T}{he} exponential proliferation of the Internet-of-things (IoT) and next-generation wireless services has generated unprecedented data volumes at the network edge.
In this evolving landscape, edge intelligence has emerged as a critical enabler, shifting the paradigm from centralized cloud processing to distributed computation \cite{EI_ProcIEEE}. 
This transition is driven by the sheer scale of privacy-sensitive information produced by billions of mobile devices and autonomous systems \cite{fediot}.
To circumvent the prohibitive latency and bandwidth costs associated with uploading raw data to centralized servers, federated learning (FL) has established itself as a robust distributed training framework \cite{FL_MEN_Survey}.
By enabling devices to train models locally and exchange only model updates, FL effectively preserves data privacy while harnessing the collective intelligence of the network.

However, the ubiquitous deployment of FL is currently impeded by an infrastructure disparity.
Prevailing FL frameworks implicitly predicate on pervasive terrestrial connectivity, yet vast regions including oceans, deserts, and remote rural areas lack such infrastructure, creating blind spots for intelligent services.
In this context, low Earth orbit (LEO) satellite networks have emerged as a paradigm-shifting solution to this coverage bottleneck \cite{satcomp}.
Distinguished by their seamless global coverage and significantly lower propagation latency compared to geostationary systems, LEO constellations are evolving into orbital edge computing platforms capable of extending the FL frontier to a planetary scale \cite{satfl_wcmag}.

Despite the promise of satellite-based FL, its practical deployment encounters distinct technical hurdles.
LEO platforms are characterized by stringent energy budgets and limited bandwidth, while their mobility induces highly dynamic topologies.
Conventional digital aggregation schemes relying on orthogonal multiple access struggle to scale in such environments and often lead to severe uplink bottlenecks \cite{FedLEO}.
To mitigate these limitations, over-the-air (OTA) computation has emerged as a promising physical-layer technique. This approach exploits the superposition property of wireless channels to aggregate analog signals simultaneously, significantly enhancing spectral efficiency \cite{fedOTA}.
To fully realize this potential, we emphasize a dual-layer OTA architecture. The first layer facilitates concurrent analog aggregation from massive ground devices to an overhead satellite. The second layer then enables multiple satellites to simultaneously forward their aggregated waveforms to a central data processing center (DPC). Furthermore, the integration of this dual-layer mechanism with beam hopping (BH) creates a powerful synergy. By enabling satellites to dynamically steer radio resources toward areas with high traffic demand, this integrated architecture can revolutionize data aggregation from space \cite{multisat}.

While dual-layer OTA aggregation in LEO satellite networks holds immense promise for achieving ubiquitous global coverage, research into such integrated learning-centric systems remains in a nascent stage.
Specifically, while foundational studies (e.g., \cite{fedOTA}) have established robust OTA-FL theories, they are largely confined to static terrestrial networks and cannot address the extreme mobility and intermittent visibility of LEO environments. To adapt FL for space, recent works have made significant strides. For instance, \cite{FedLEO} proposed a decentralized FedLEO framework, and \cite{satfed2} investigated asynchronous OTA-FL using high-altitude platforms (HAPs) for space-to-air aggregation. However, \cite{FedLEO} still relies on traditional digital access, where latency scales linearly with device density, and \cite{satfed2} primarily treats satellites as clients training on local remote sensing data, fundamentally not addressing the massive ground-to-satellite connectivity bottleneck under strict multi-beam payload constraints. In contrast, our work provides a novel cross-layer orchestration that explicitly binds the physical-layer OTA aggregation constraints with the spatial-temporal scheduling of adaptive BH. The core system-level insight is that by dynamically illuminating ground cells based on real-time data freshness, the LEO satellite can proactively optimize spatial resource efficiency and completely decouple aggregation latency from massive IoT device density.
This article provides a comprehensive overview of dual-layer OTA-enabled FL in LEO satellite networks, covering the system architecture, key advantages, application scenarios, and a proximal policy optimization (PPO)-based case study, followed by a discussion of remaining challenges and future directions.

\section{Dual-Layer Architecture Design}
\label{sec:sys}
This section first introduces the OTA aggregation principle, followed by the adaptive BH mechanism and the overall system workflow.
\subsection{Over-the-Air Computation}

Conventional wireless systems are designed to suppress interference and treat superposed signals as data corruption, an interference-avoidance paradigm that imposes a severe bottleneck for massive satellite IoT \cite{guoAnalogAgg}. Standard orthogonal digital access schemes such as time division multiple access (TDMA) or orthogonal frequency division multiple access (OFDMA) allocate exclusive resource blocks to individual devices, so aggregation latency scales linearly with the number of participants, rendering the timely collection of updates from thousands of ground sensors practically infeasible within the constrained LEO visibility windows \cite{satfed2}.

OTA represents a fundamental paradigm shift by harnessing interference as a constructive mechanism \cite{comudo}. It leverages the natural superposition property of electromagnetic waves, allowing all participating ground devices to transmit analog model parameters simultaneously over the shared frequency band. The wireless channel then performs a weighted summation directly at the receiver antenna, which aligns intrinsically with the averaging operation required by FL algorithms.
The architecture realizes this concept through a robust dual-layer protocol.
In the first layer ground devices execute channel inversion power control to align their signal magnitudes at the satellite receiver and create a coherent superposition representing the aggregated local model. While effective for achieving coherent aggregation, this mechanism imposes an energy trade-off. IoT devices experiencing deep fading must significantly scale up their transmit power, which can accelerate battery depletion in energy-constrained deployments. In practice, strict channel inversion can be complemented by more energy-aware strategies such as truncated channel inversion, adaptive power capping, or partial device participation. These approaches intentionally relax perfect signal alignment for devices with poor channel conditions, thereby balancing aggregation fidelity against long-term device longevity.
Crucially the satellite does not attempt to decode individual digital messages but functions as a transparent analog relay.
In the second layer, it amplifies and forwards this noisy, aggregated waveform to the ground DPC.
It is important to note that the second-layer OTA aggregation introduces distinct physical-layer challenges compared to the ground-to-satellite link. To handle the severe differential propagation delays and significant Doppler shifts across multiple satellites simultaneously transmitting to the DPC, the framework relies on the advanced hardware capabilities of space platforms. Utilizing highly predictable ephemeris data and precise onboard clocks, satellites perform deterministic timing advance and Doppler pre-compensation to ensure that signals arrive at the DPC phase-aligned. Furthermore, the DPC is assumed to be a high-performance gateway equipped with a high-gain phased array antenna to robustly process the superposed analog waveforms. The viability of this space-to-ground OTA paradigm has been strongly supported by recent studies \cite{satfed2}.
This mechanism effectively decouples communication latency from the user population size and enables the system to support massive connectivity with constant latency overhead.

However, this efficiency comes with trade-offs. Unlike digital schemes that rely on robust coding to correct errors, OTA is sensitive to channel noise and requires precise synchronization. In the dual-layer architecture, the end-to-end aggregation mean squared error (MSE) decomposes into three additive terms. The first is a bias term from the mismatch between the analog combining weights (set by the channels and power coefficients) and the desired digital aggregation weights, which persists even without noise because analog superposition cannot perfectly realize the target weighted average. The second is the uplink noise re-scaled across both hops, whose impact on the global model depends on the satellite-relay channel and the normalization factors. The third is the ground-side noise at the DPC, which sets a fundamental noise floor independent of the uplink. Since the satellite is a transparent analog relay, the uplink disturbance compounds in the second hop, yielding the cascading distortion. Keeping this total distortion below the threshold penalized in the scheduling reward preserves FL convergence \cite{fedOTA}, though holding it under severely degraded channels is increasingly hard. In practice, ephemeris-based Doppler pre-compensation is inevitably imperfect. Residual Doppler and phase errors manifest as a slow phase rotation that perturbs the effective analog combining weights, enlarging the bias term above rather than causing unbounded distortion. Being small under deterministic orbital prediction and further constrained by the MSE-aware scheduling reward, their impact on the aggregated global model stays bounded. Table \ref{tab:comparison} provides a qualitative comparison of this analog approach against established digital schemes, specifically orthogonal multiple access (OMA) and non-orthogonal multiple access (NOMA), highlighting the distinct operational characteristics of each supported by the analyses in \cite{WFL_DigAna}.

\begin{table}[ht]
\caption{Comparison of Physical Layer Aggregation Schemes}
\label{tab:comparison}
\centering
\begin{tabular}{m{0.195\linewidth} m{0.236\linewidth} m{0.18\linewidth} m{0.2\linewidth}}
\hline
\textbf{Feature} & \textbf{Dual-Layer OTA} & \textbf{Digital OMA} & \textbf{Digital NOMA} \\
\hline
\textbf{Access Mechanism} & Waveform Superposition & Time/Freq Division & Power Domain Superposition \\
\hline
\textbf{Bandwidth Efficiency} & High & Low & Moderate \\
\hline
\textbf{Access Latency} & Constant & Linear Growth & User-dependent \\
\hline
\textbf{Receiver Complexity} & Low & Moderate & High \\
\hline
\textbf{Noise Robustness} & Low & High & Moderate \\
\hline
\textbf{Hardware Design} & Simple & Standard & Complex \\
\hline
\textbf{Application Suitability} & Massive-scale model aggregation & Small-scale & Medium-scale \\
\hline
\end{tabular}
\end{table}

\subsection{Adaptive Beam Hopping}

While OTA computation solves the uplink capacity issue, the spatial distribution of resources presents another challenge. LEO satellites move at high velocities relative to the Earth, causing the ground topology to change rapidly. Traditional satellites use fixed multi-beam antennas that continuously illuminate a static grid of cells, wasting power on regions with no active devices or stale data while under-serving areas with urgent model updates \cite{channel}.

To address this spatial inefficiency, the architecture incorporates adaptive BH technology governed by a deep reinforcement learning (DRL) agent. 
This intelligent mechanism allows the satellite to dynamically activate a limited subset of spot beams at any given time slot, effectively steering radio resources to where they are most needed \cite{BH_DRL2}. The DRL agent learns illumination decisions driven by the freshness of the training data accumulated by ground devices and the instantaneous channel conditions, prioritizing areas with significant new data generation while skipping regions with outdated or sparse data. This on-demand strategy ensures that the limited onboard energy is focused exclusively on capturing high-value model updates, allowing the system to adapt in real-time to the uneven distribution of users and the orbital movement of the constellation \cite{BH2}.

To formally quantify data freshness in alignment with the standard age of information (AoI) framework, we model it as an AoI-discounted effective data volume. As local observations age, previously cached data loses relevance. Mathematically, the effective data volume for device $k$ at round $t$ is
$\mathcal{D}_{k}(t) = \min \Big( D_{\max}, \bar{\mathcal{D}}_{k}(t) + (1 - a_{k}(t-1)) \eta \mathcal{D}_{k}(t-1) \Big),$
where $\bar{\mathcal{D}}_{k}(t)$ is newly generated data, $D_{\max}$ is the buffer capacity, $a_k(t-1) \in \{0,1\}$ is the previous scheduling indicator, and $\eta \in (0,1)$ is a freshness discount factor. This ensures stale data with a high AoI is intrinsically penalized.

To rigorously ground this intelligent orchestration without delving into heavy mathematical derivations, we formalize the decision-making process as a Markov decision process (MDP) defined by three core components:
1) \textbf{State}: At each scheduling slot, the DRL agent observes the global network environment. This includes the pending data queues across devices to reflect data freshness, alongside the real-time channel conditions for both the device-to-satellite and satellite-to-DPC links.
2) \textbf{Action}: The agent outputs a joint resource allocation policy comprising the discrete beam activation vectors for each satellite, alongside the continuous transmit power coefficients for the participating devices.
3) \textbf{Reward}: The agent receives a scalar reward designed to maximize the volume of successfully aggregated data. This reward is reduced by a bounded penalty term whenever the physical-layer OTA aggregation distortion, quantified by the mean squared error, exceeds a strict predefined threshold.

\subsection{Overall System Architecture and Operational Flow}

\begin{figure}[t]%
			\centering
			\includegraphics[width=0.49\textwidth]{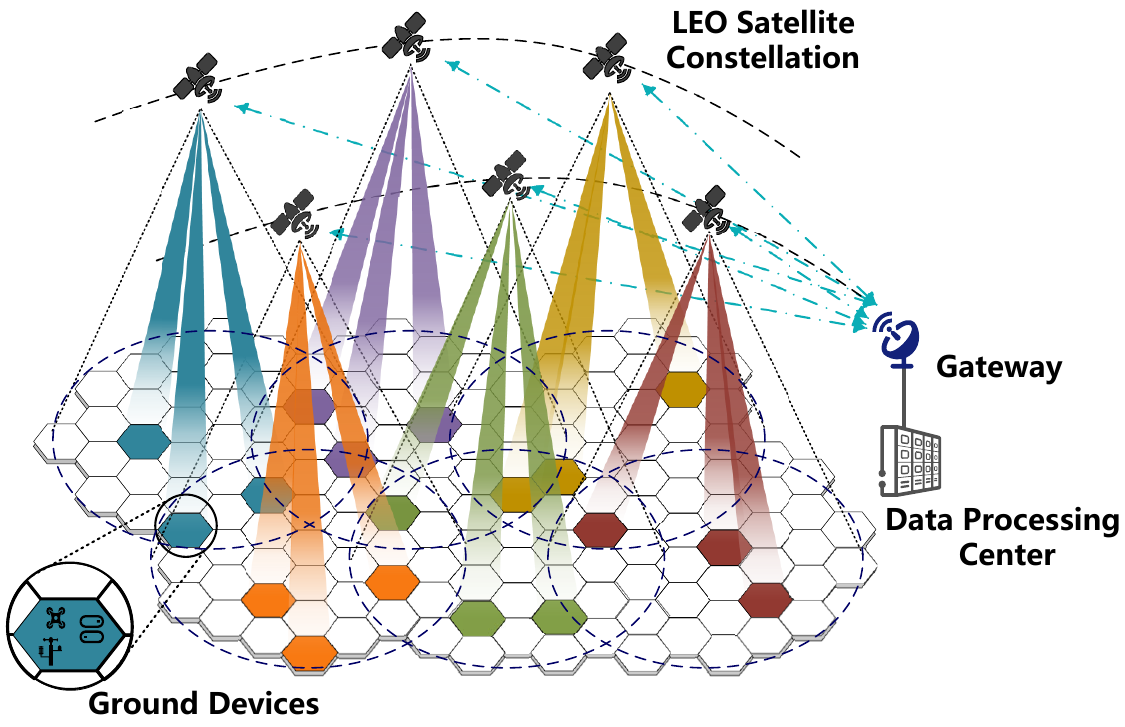}
			\caption{BH framework for dual-layer OTA-enabled FL in LEO satellite networks.}
			\label{fig:system model}
		\end{figure}

The complete system integrates these two technologies into a cohesive, closed-loop learning environment involving ground devices, LEO satellites, and a central DPC. As illustrated in Fig. \ref{fig:system model}, the architecture relies on a hierarchical structure where satellites serve as intelligent relays between the scattered IoT devices and the DPC.

\begin{figure}[t]
    \centering
    \includegraphics[width=0.49\textwidth]{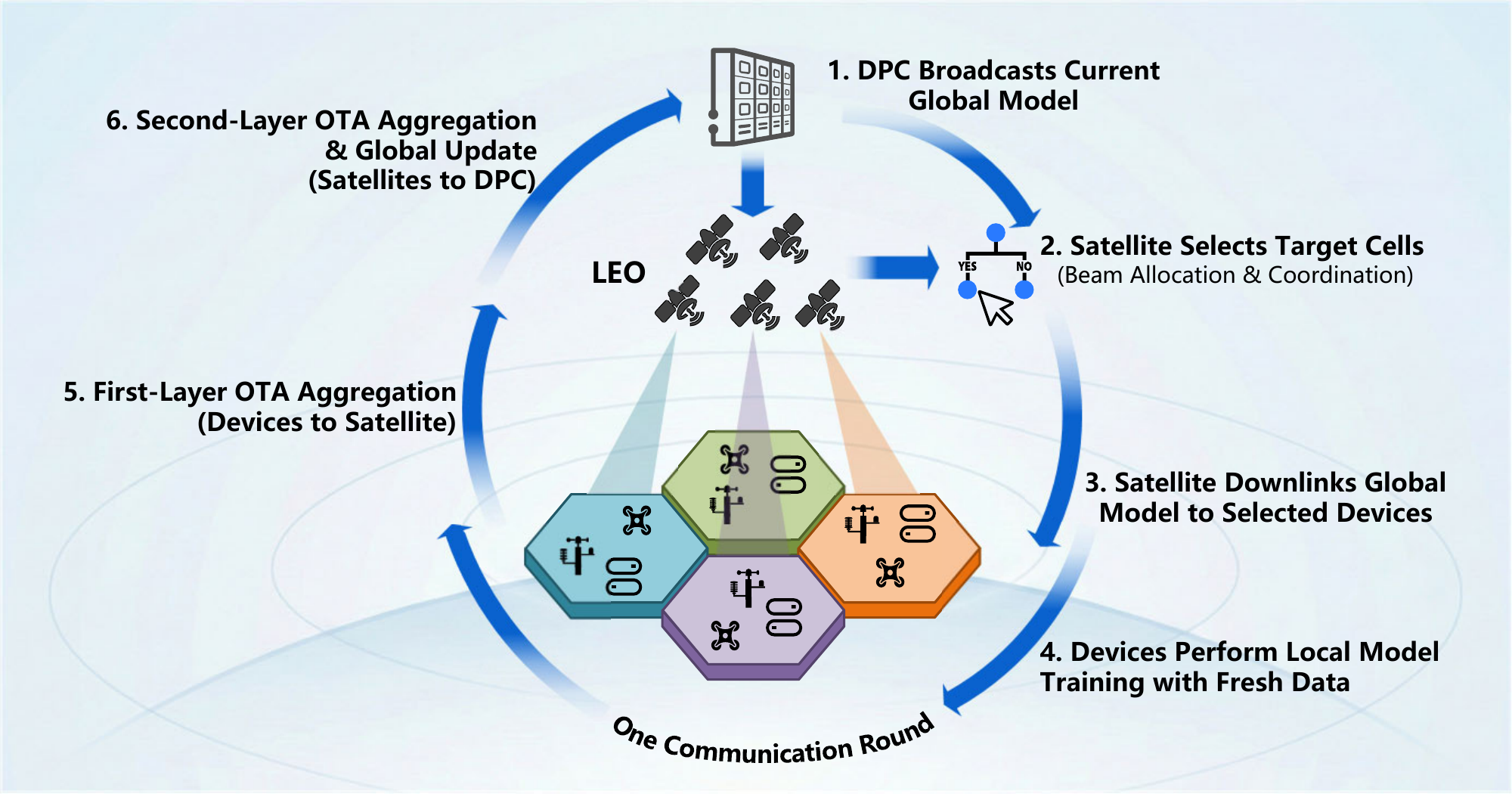} 
    \caption{Operational workflow of the dual-layer OTA FL process.}
    \label{fig:workflow}
\end{figure}

To illustrate the operational workflow, Fig. \ref{fig:workflow} details the step-by-step learning process. 
Operating centrally at the DPC, the DRL agent jointly coordinates the BH patterns across multiple satellites.
\textcircled{\small 1} At the beginning of each round, the DPC broadcasts the current global model parameters and the coordinated BH schedules to the LEO satellite constellation. \textcircled{\small 2} Upon receiving the model, satellites execute a downlink BH pattern to illuminate specific target cells, which are selected based on data freshness and channel conditions, thereby distributing the global model to the ground users.
\textcircled{\small 3} Subsequently, active IoT devices utilize these parameters to perform local training on their newly collected environmental data. \textcircled{\small 4} Once local model updates are computed, devices perform the first layer of OTA transmission, sending their updates simultaneously. \textcircled{\small 5} The satellite captures the superimposed analog signal and immediately executes the second layer of transmission, forwarding the aggregated waveform to the DPC. \textcircled{\small 6} Finally, the DPC receives the superposition of signals from multiple satellites, completes the global aggregation, and initiates the next training round. 
This cyclical process leverages the high bandwidth efficiency of dual-layer OTA to handle massive device access and the flexibility of BH to manage the dynamic satellite-ground topology, ensuring rapid and energy-efficient model convergence.

A critical advantage of this learning-centric architecture is its compatibility with resource-constrained space platforms. The computationally intensive DRL training and inference processes are executed entirely on the ground-based DPC. The LEO satellites merely receive and execute the lightweight scheduled action tables uploaded during their visibility windows, resulting in negligible onboard computational overhead. Furthermore, by adopting PPO, the agent leverages a clipped surrogate objective that restricts the step size of policy updates. Along with the bounded reward design, this mechanism prevents catastrophic performance drops and ensures remarkable training stability even within highly dynamic satellite environments.
Regarding the feedback loop latency, since the DRL agent operates centrally at the DPC, there exists an inherent delay between CSI collection, policy computation, and action execution aboard the fast-moving satellite. To mitigate this, the scheduler leverages deterministic ephemeris data to predict future channel conditions and topology evolution, and pre-computes scheduling action tables that are uploaded to the satellite before execution. Furthermore, the duration of each FL round is significantly shorter than the typical satellite visibility window (on the order of minutes), allowing multiple decision epochs to be refreshed within a single satellite pass. To further guard against command obsolescence, namely the actual state deviating from the pre-computed policy within the execution window, the scheduler exploits the fact that LEO dynamics are dominated by deterministic orbital geometry. Meanwhile, residual stochastic fluctuations are absorbed by the MSE-penalized reward, keeping pre-computed actions robust to bounded deviations. Upon a significant mismatch, lightweight state feedback lets the DPC refresh the action table within the same pass.

\section{Advantages of the Proposed Architecture}
The integration of dual-layer OTA computation with adaptive BH creates a synergy that resolves the conflict between the resource scarcity of LEO satellites and the massive connectivity requirements of global FL. By shifting from a communication-centric to a learning-centric design, this architecture offers three distinct advantages for non-terrestrial networks (NTN).
\subsection{Ultra-Scalable Low-Latency Aggregation}

The primary advantage of the dual-layer OTA architecture is its ability to decouple aggregation latency from ground device density. In conventional satellite links, thousands of IoT devices compete for scarce time/frequency resources, creating a linear uplink bottleneck that often exceeds the short visibility window. By enabling simultaneous analog transmissions and in-orbit aggregation, our framework compresses the uplink into a single resource block: aggregation time is constant whether ten or ten thousand devices participate, transforming the satellite into an efficient on-orbit aggregator.

\subsection{Spatio-Temporal Resource Efficiency}

Fixed beam coverage patterns waste precious on-board power and spectrum when satellites traverse sparsely populated or low-value regions. 
To address this issue, the proposed adaptive BH mechanism uses a DRL-based scheduler to dynamically allocate beam resources according to data freshness and channel conditions. 
This on-demand illumination minimizes energy spent on idle or stale regions and concentrates payload resources where they most improve learning performance, ensuring an operationally efficient LEO payload that sustains high-quality online FL under tight budgets.

\subsection{Privacy, Robustness and Learning-Centric Fidelity}

Beyond throughput and agility, the architecture yields intrinsic privacy and robustness benefits. OTA aggregation naturally mixes signals in the channel such that the receiver observes only a noisy weighted sum rather than distinct per-device gradients. This physical-layer blending complicates model-inversion or packet-level interception attacks and effectively complements the algorithmic privacy of FL. Furthermore, the framework explicitly balances data quantity against aggregation distortion. By regulating transmit powers and beam selection based on channel quality and data freshness, the system prevents noisy updates from degrading the global model. This learning-centric optimization promotes faster convergence and higher final accuracy compared to greedy or throughput-first strategies, delivering reliable intelligence despite the harsh and variable satellite channel.

\section{Application Scenarios}
\label{sec:app}
\begin{figure*}[t]%
			\centering
			\includegraphics[width=0.98\textwidth]{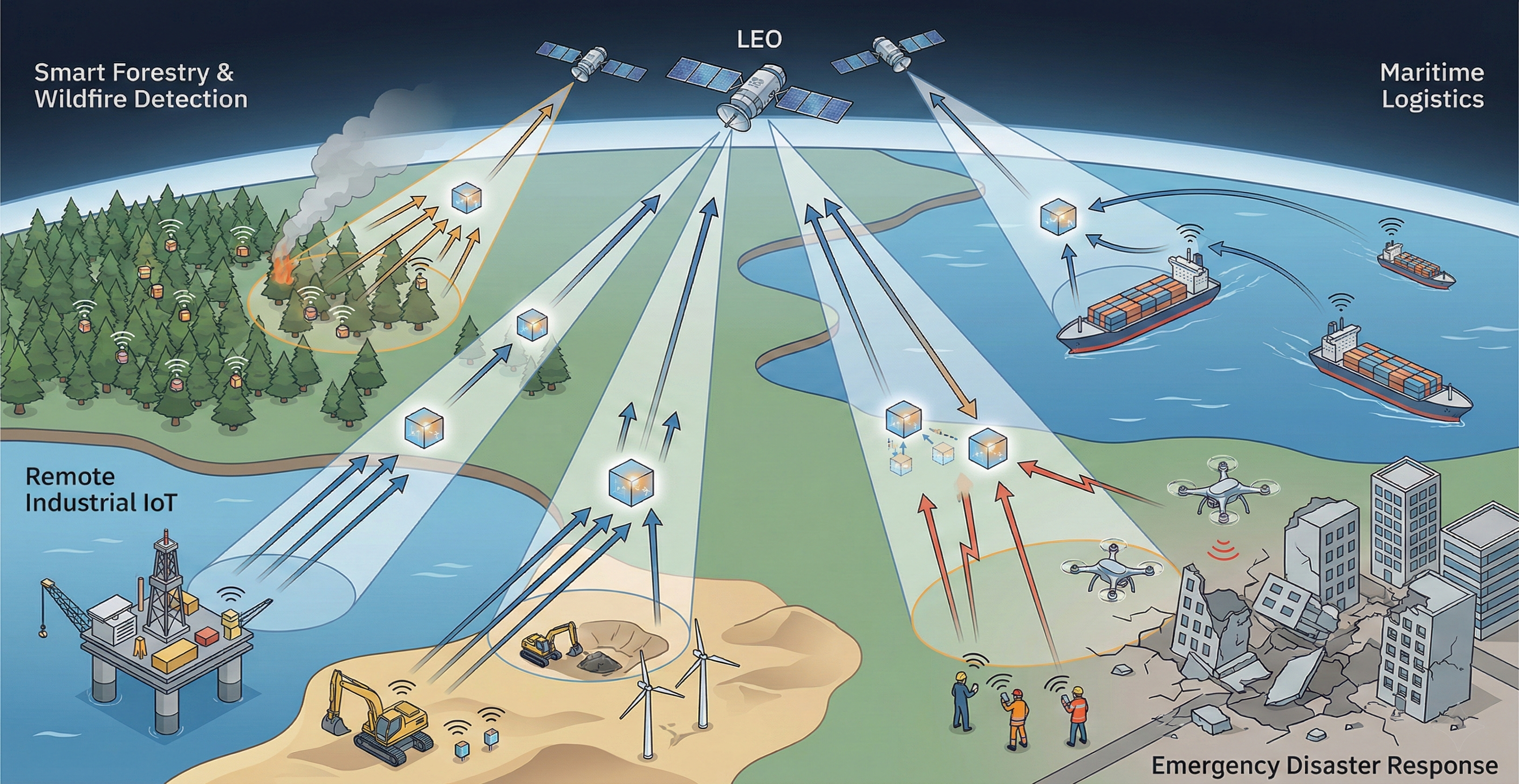}
			\caption{Illustration of key application scenarios for OTA-enabled FL in LEO satellite networks.}
\label{fig:applications}
		\end{figure*}

By transforming LEO satellites into edge aggregation nodes, this framework unlocks new possibilities across several vertical industries ranging from smart forestry and wildfire detection to emergency disaster response, as shown in Fig. \ref{fig:applications}.

\subsection{Smart Forestry and Wildfire Detection}

A critical application is the early detection of wildfires across vast, uninhabited forests, where thousands of low-power IoT sensors monitor temperature, humidity, and smoke levels. The OTA-enabled architecture allows these sensors to locally train lightweight anomaly detection models and upload only model updates simultaneously during a satellite pass, while freshness-aware scheduling ensures the aggregated global model reflects the current environmental status for rapid detection of ignition events.

\subsection{Maritime Logistics}

The maritime industry operates primarily in international waters beyond cellular coverage, where modern vessels carry sophisticated sensors for predictive maintenance, route optimization, and cargo monitoring. Since shipping companies are reluctant to share commercially sensitive raw data, FL trains collective industry models without exposing proprietary information, while adaptive BH dynamically illuminates vessel clusters to aggregate their local models into a collaborative intelligence network across the high seas.

\subsection{Remote Industrial IoT}

Remote industrial operations, such as offshore oil rigs, mining sites, and distant wind farms, generate massive telemetry data yet rely on expensive, bandwidth-constrained satellite backhaul. Local training with dual-layer OTA aggregation removes this heavy uplink burden; for instance, geographically dispersed wind turbines can collaboratively learn to predict mechanical failures from vibration patterns, with the LEO constellation acting as the global coordinator.

\subsection{Emergency Disaster Response}

In the aftermath of catastrophes such as earthquakes or tsunamis, terrestrial infrastructure is often compromised, and LEO satellites immune to ground damage become the primary communication backbone. Aggregating information from survivor devices and rescue drones, the low-latency OTA process trains a global model that builds a real-time disaster heatmap and identifies high-priority rescue zones, while prioritizing fresh data keeps rescue efforts directed by the current situation.

\section{Case Study}

We conducted a comprehensive evaluation within a simulated IoT deployment served by a 550 km altitude LEO constellation. To demonstrate the system's potential, we implemented the learning-aware scheduler using PPO and compared its system-level performance against alternative DRL-based scheduling policies (DDPG, TD3, and SAC) as well as a non-learning greedy heuristic that allocates beams to the cells with the best instantaneous channel conditions. Our evaluation focuses on two architectural goals: global model learning efficiency and spatial resource utilization.
Compared with a standard digital FL benchmark such as satellite FedAvg over orthogonal links, the gain is quantifiable: under an identical bandwidth budget, the digital baseline serializes device uploads so that its per-round uplink airtime grows linearly with the device count, whereas dual-layer OTA completes each aggregation within a single resource block. Within a fixed visibility window, the framework thus sustains roughly $K$-fold more aggregation rounds for $K$ devices per cell, translating into fewer required satellite passes and higher bandwidth efficiency, consistent with Table~\ref{tab:comparison}.


For reproducibility, key simulation parameters include: 3 devices per illuminated cell, a 20 GHz carrier frequency, free-space path loss governed by dynamic slant distance, and a 354.81 K receiver noise temperature. To maintain stable OTA analog aggregation, ground terminals utilize ephemeris-based synchronization techniques to actively pre-compensate for the deterministic LEO Doppler shifts prior to transmission.

\subsection{Accelerated Learning Convergence}
\begin{figure}[t]
\centering
\includegraphics[width=0.48\textwidth]{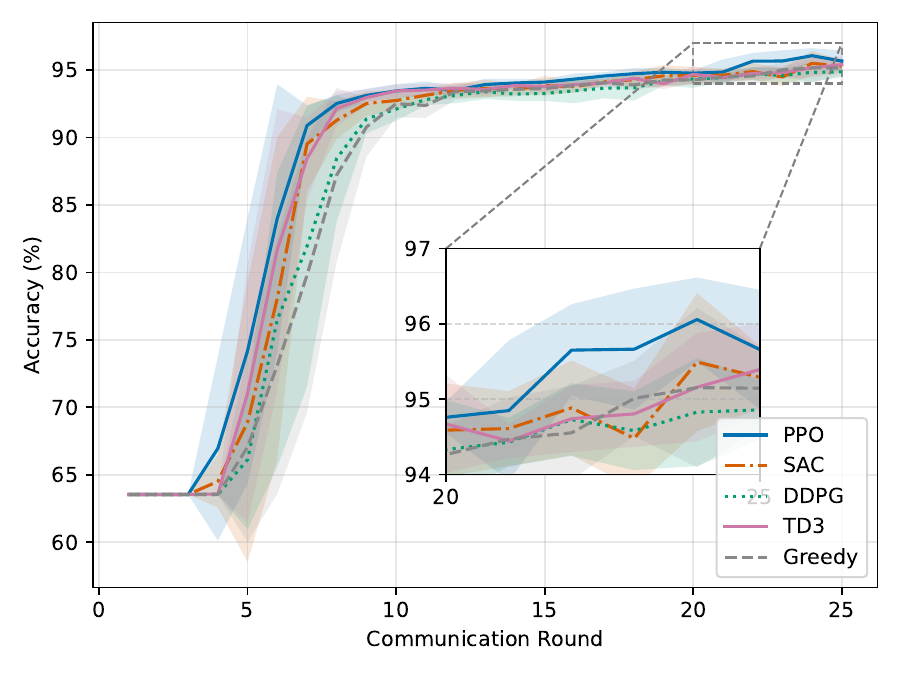} 
\caption{Evolution of global test accuracy versus communication rounds.}
\label{fig:accuracy}
\end{figure}

The primary goal of edge intelligence is to achieve a high-accuracy model with minimal communication rounds. As illustrated in Fig. \ref{fig:accuracy}, we tracked the test accuracy of a convolutional neural network (CNN) training on the FLAME (Fire Luminosity Airborne-based Machine learning Evaluation) dataset, which explicitly aligns our evaluation with the wildfire detection application discussed in Section \ref{sec:app}. 


While the DRL baselines (DDPG, TD3, and SAC) demonstrate the ability to adapt to system dynamics, their learning behaviors vary due to the complexity of the hybrid action space. The greedy heuristic, which selects cells solely based on instantaneous channel conditions without considering data freshness or long-term planning, is comparable to DDPG and TD3 in some communication rounds, but does not consistently match the accuracy of PPO and SAC. 
Notably, although advanced baselines like TD3 and SAC show highly competitive convergence on this specialized dataset, the proposed PPO agent ultimately achieves and maintains the highest final test accuracy. By explicitly balancing channel conditions with data freshness, PPO ensures robust policy updates, yielding reliable intelligence which is vital for time-sensitive applications like wildfire monitoring.

\subsection{Resource Utilization Under Constraints}

\begin{figure}[t]
\centering
\includegraphics[width=0.48\textwidth]{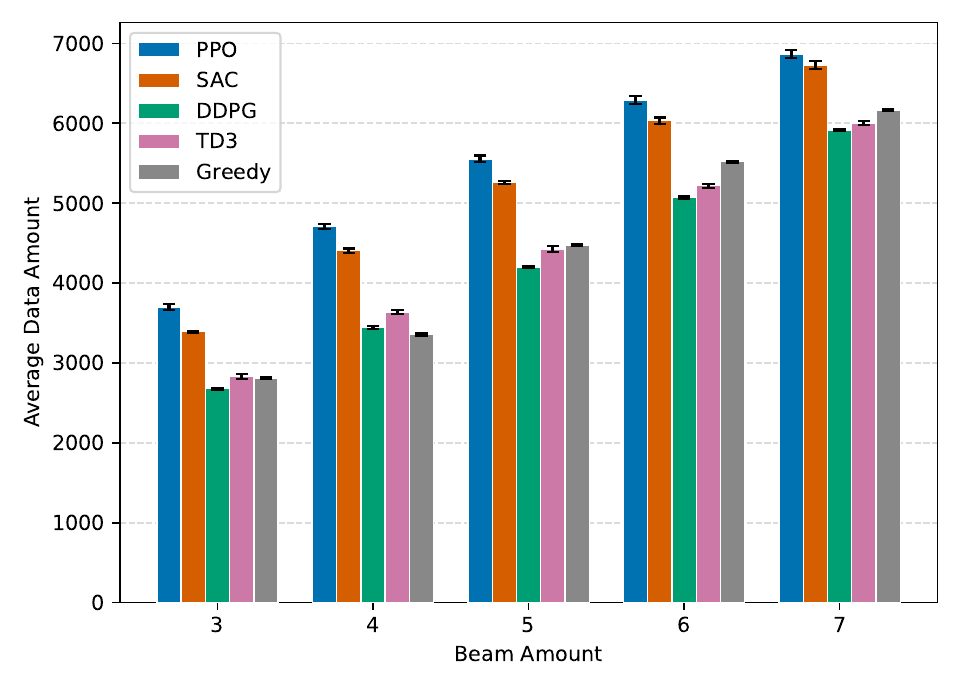}
\caption{Average training data collected per round versus the maximum number of beams per satellite.}
\label{fig:beams}
\end{figure}

We investigated how effectively different strategies utilize the stringently limited simultaneous active spot beams of LEO satellites to capture valuable data. 
Fig. \ref{fig:beams} presents a comparative bar chart of average effective data volume collected per round, with error bars explicitly indicating the standard deviation across the five independent runs.

The bar chart reveals a significant efficiency gap across the algorithms. While increasing the beam budget naturally allows for greater data aggregation, the height difference between the bars highlights the superior resource orchestration of the proposed method. The baseline adaptive algorithms, namely SAC, TD3, DDPG, and the greedy heuristic, show competitive utilization levels but still fall short of the optimal capacity. In contrast, even when accounting for training variance (i.e., comparing the lower bound of PPO's performance against the upper bounds of the baselines), the proposed PPO approach consistently outperforms these benchmarks across every configuration. Notably, the intelligent scheduler demonstrates superior efficiency even with limited resources, indicating that the optimized software-defined policy can effectively mitigate hardware limitations without requiring excessive beam resources.

\section{Challenges and Future Directions}

While the dual-layer OTA framework significantly advances satellite-based edge intelligence, the evolution of NTN presents new opportunities, and future research should focus on several key areas to further robustify and scale this paradigm.

\subsection{Cooperative Multi-Satellite Aggregation}
Current architectures typically treat satellites as isolated aggregation nodes. However, the rapid deployment of mega-constellations will increasingly result in scenarios where ground devices fall within the overlapping coverage footprints of multiple satellites simultaneously. Future architectures should move beyond single-satellite operations to explore cooperative swarm intelligence where satellites dynamically form temporary aggregation clusters via inter-satellite links. By distributing the computational load across a mesh of satellites, the network can support significantly larger model sizes. Furthermore, this cooperative approach enables distributed consensus mechanisms that improve resilience against single-point failures. Importantly, the ground-heavy DRL scheduler can naturally coordinate inter-satellite handover boundaries by predicting visibility windows via ephemeris data, ensuring that FL aggregation is not interrupted when ground cells migrate between adjacent satellites' footprints.

\subsection{Heterogeneous Model and Data Architectures}
In realistic large-scale IoT deployments, the participating ground devices possess widely varying hardware capabilities that range from powerful edge servers to energy-constrained passive sensors. Imposing a uniform model architecture across all nodes is often impractical and excludes low-power devices from contributing valuable data. 
Future research is needed to adapt analog aggregation schemes for heterogeneous networks where devices may transmit updates for different sub-models or compressed model variants. Notably, the dual-layer OTA architecture can naturally accommodate heterogeneous learning paradigms by integrating lightweight sub-models or compressed model updates tailored to device capabilities. Existing techniques such as model pruning, knowledge distillation, and heterogeneous FL can be incorporated to enable resource-constrained devices to participate in collaborative learning while reducing their computational and energy burdens. As long as the participating devices project their updates onto a shared subspace for aggregation, the analog superposition property is preserved even under heterogeneous model sizes.
In addition, practical IoT deployments exhibit significant data heterogeneity across spatially separated clusters. While the proposed PPO scheduler does not rely on the IID assumption, since it optimizes scheduling policies according to the system reward rather than local model gradients, severe non-IID distributions primarily affect the FL convergence process. In such scenarios, fresher data are not necessarily more statistically representative. Future scheduling policies should jointly account for data freshness and statistical diversity, for example by incorporating data heterogeneity indicators into the scheduling state or reward function.

\subsection{Robustness Against Imperfect CSI and Environmental Uncertainties}
The performance of OTA computation relies heavily on precise CSI to perform accurate power control and phase alignment during signal superposition. 
In high-mobility LEO environments where satellites travel at velocities exceeding 7 km/s, obtaining perfect real-time CSI is extremely challenging due to significant Doppler shifts and propagation delays. Future work must develop robust aggregation schemes that can tolerate inevitable channel estimation errors without causing catastrophic model divergence. 
Potential solutions include blind or semi-blind aggregation techniques that eliminate strict phase synchronization requirements, or AI-driven channel prediction modules deployed onboard to preemptively adjust transmission parameters. Beyond CSI imperfections, atmospheric impairments prevalent at Ka/Ku bands, including tropospheric scintillation and rain fading, introduce additional propagation losses that reduce the received SNR. This may degrade the accuracy of channel estimation and channel inversion power control, resulting in imperfect signal alignment during OTA aggregation and increased aggregation error. Incorporating atmospheric channel prediction, adaptive power control, or weather-aware beam scheduling into the DRL agent represents a promising direction to mitigate these effects.

\section{Conclusion}

This article has outlined an adaptive BH dual-layer OTA FL framework for LEO satellite networks. Compared with traditional digital access schemes and fixed-beam systems, this architecture leverages the high bandwidth efficiency of OTA aggregation and the spatial flexibility of BH, significantly improving uplink spectrum efficiency while decoupling aggregation latency from device density for agile and learning-centric resource allocation.
Moreover, the case study confirms that the proposed PPO scheduling policy accelerates global model convergence and optimizes data utilization. These advantages enable the system to adapt to dynamic non-terrestrial environments and meet the massive connectivity demands of future global IoT applications. Since research on OTA-enabled satellite edge intelligence is still in its infancy, we hope this article provides valuable references for future theoretical research and practical implementations.

\bibliographystyle{IEEEtran}
\bibliography{reference}
 
\end{document}